\documentclass[aps, preprint, onecolumn, pre,superscriptaddress,amsmath,amssymb]{revtex4-1}

\usepackage{color}      % use if color is used in text
\usepackage{amsmath}

\usepackage[]{graphicx}

\usepackage{amssymb,amsfonts,amsmath}
\usepackage{mathrsfs}

\newcommand{\fref}[1]{Fig.~\ref{fig:#1}}

\newcommand{\flabel}[1]{\label{fig:#1}}
\newcommand{\eref}[1]{Eq.~\ref{eqn:#1}}

\newcommand{\elabel}[1]{\label{eqn:#1}}

\newcommand{\beq}{\begin{equation}}
\newcommand{\beqs}{\begin{subequations}}
\newcommand{\beqnn}{\begin{equation*}}
\newcommand{\beqn}{\begin{eqnarray}}
\newcommand{\beqnnn}{\begin{eqnarray*}}
\newcommand{\bml}{\begin{multline}}
\newcommand{\eeq}{\end{equation}}
\newcommand{\eeqs}{\end{subequations}}
\newcommand{\eeqnn}{\end{equation*}}
\newcommand{\eeqn}{\end{eqnarray}}
\newcommand{\eeqnnn}{\end{eqnarray*}}
\newcommand{\eml}{\end{multline}}

\newcommand{\avg}[1]{\langle{#1}\rangle}

\newcommand{\td}{\ensuremath{T_{\mathrm{d}}}}
\newcommand{\tpeak}{\ensuremath{T_\mathrm{PtP}}}
\newcommand{\mudil}{\ensuremath{\mu_{\mathrm{d}}}}

\begin{document}

\title{Discrete gene replication events drive coupling between the cell cycle and circadian clocks}

\author{Joris Paijmans}
\affiliation{FOM Institute AMOLF, Science Park 104, 1098 XG Amsterdam, The Netherlands}
\author{Mark Bosman}
\affiliation{FOM Institute AMOLF, Science Park 104, 1098 XG Amsterdam, The Netherlands}
\author{Pieter Rein ten Wolde}
\affiliation{FOM Institute AMOLF, Science Park 104, 1098 XG Amsterdam, The Netherlands}
\author{David K. Lubensky} 
\affiliation{Department of Physics, University of Michigan, Ann Arbor, MI 48109-1040},

\begin{abstract} 
  Many organisms possess both a cell cycle to control DNA
  replication and a circadian clock to anticipate changes between
  day and night. In some cases, these two rhythmic systems are known
  to be coupled by specific, cross-regulatory interactions.  Here, we
  use mathematical modeling to show that, additionally, the cell cycle
  generically influences circadian clocks in a non-specific fashion:
  The regular, discrete jumps in gene-copy number arising from DNA
  replication during the cell cycle cause a periodic driving of the
  circadian clock, which can dramatically alter its behavior and
    impair its function. A clock built on negative transcriptional
  feedback either phase locks to the cell cycle, so that the
  clock period tracks the cell division time, or 
   exhibits erratic behavior. We argue that the cyanobacterium 
   {\em Synechococcus elongatus} has evolved two features that protect 
   its clock from such disturbances, 
   both of which are needed to fully insulate it
  from the cell cycle and give it its observed robustness: a
phosphorylation-based protein modification oscillator,
together with its accompanying push-pull read-out circuit that
  responds primarily to the ratios of the different phosphoforms,
makes the clock less susceptible to perturbations in
    protein synthesis; and the presence of
  multiple, asynchronously replicating copies of the same chromosome
  diminishes the effect of replicating any single copy of a gene.
   
   \vspace*{12pt}
   \noindent {\bf Significance Statement}: Huygens famously showed that
   two mechanically connected clocks tend to tick in synchrony.  We
   uncovered a generic mechanism that can similarly phase lock two
   rhythmic systems present in many living cells: the cell cycle and
   the circadian clock.  DNA replication during the cell cycle causes
   protein synthesis rates to show sharp, periodic jumps that can
   entrain the clock.  To faithfully keep time in the face of these
   disturbances, circadian clocks must incorporate specific insulating
   mechanisms.  We argue that, in cyanobacteria, the presence of
   multiple, identical chromosome copies and the clock's core
   protein-modification oscillator together play this role.  Our
   results shed new light on the complex factors that constrain the
   design of biological clocks.
\end{abstract}

\maketitle

\section{Introduction}
Circadian clocks---autonomous oscillators with a roughly 24 hour
period that can be entrained to daily cycles of light and dark ---are thought to
confer important advantages on living cells by allowing them to anticipate diurnal
environmental changes.  Recent decades have seen considerable progress
in elucidating both the architecture and the function of these
biological timekeepers.  Circadian clocks, however, are not the only
oscillatory systems present in living cells.  Most notably, cell
growth and division are governed by a cell cycle, which can in many contexts be viewed as an autonomous oscillator. Much recent attention has been directed towards the connections between
these two rhythmic systems, which are relevant for processes ranging from plants' response to
shade~\cite{Salter2003} to cancer susceptibility
\cite{Johnson2010,Sotak2014}.  In particular, it is now
clear that circadian clocks can exert specific regulatory influences
on the cell cycle, and a number of experimental and modeling studies
have sought to tease out the implications of this regulation
\cite{Mori1996,Matsuo2003,Nagoshi2004,Dong2010,Yang:2010ci,Gerard2012,Feillet2014,Bieler2014}. Here,
we argue that, in addition to direct, specific regulation of one oscillator
by the other, there must also be more generic connections between the
circadian clock and the cell cycle
\cite{Johnson2010,ThesisBosman,Feillet2014,Bieler2014}.  In particular, we focus on
the consequences of the discrete gene replication events that accompany DNA replication.  We show that, as a result of the regular
jumps in gene copy number caused by these events, the cell cycle must,
very generally, contribute a periodic forcing to the circadian clock.  This forcing can markedly change clock behavior and degrade clock function.  We propose that cyanobacterial clocks have evolved specific features that can mitigate this effect. More broadly, this generically strong coupling to the cell cycle implies important constraints on the design of biological timekeepers if they are to remain accurate in dividing cells.

It is widely accepted that protein levels depend on a cell's gene
dosage.  Typically, a doubling of the number of chromosomal copies of
a gene should lead to an approximate doubling of its mRNA synthesis
rate and thus to a corresponding increase in its protein levels.  Most
often, however, such effects are considered in the context of a change
in the number of autosomal gene copies that persists throughout an
organism's lifetime~\cite{Volfson2006}, as, e.g., in the
haploinsufficiency of certain genes~\cite{Irish1987}.  It is less
often acknowledged that the number of copies of all genes varies over
each cell cycle, despite evidence that these variations have
measurable consequences
\cite{Trcek2011,Zopf:2013bb,Narula:2015fw,Hensel:2015vd}.  Because of
the well-known phenomenon of phase locking of oscillators
\cite{Pikovsky2003}, regular, periodic changes in gene dose are likely
to be especially relevant to cellular oscillators that depend on gene
expression. A circadian clock that became slaved to the cell
  cycle, for example, would lose its identity as an autonomous
  timekeeper, and thus much of its ability to perform its biological
  function.  Here, we show that oscillators built on negative
  transcriptional feedback---a common motif in both prokaryotic
  and eukaryotic clocks---are indeed very strongly affected by driving
  from periodic gene replication events.  This immediately raises the
  question of how real biological clocks are able to
  function in growing, dividing cells.  To address this, we
  study the circadian clock of the cyanobacterium
  \textit{Synechococcus elongatus}, which is known to exhibit
  stable rhythms over a wide range of growth rates
\cite{Mori:2001fn,Teng2013}, but whose clock appears not to
  regulate DNA replication~\cite{Mori1996}, suggesting exactly
  the sort of unidirectional forcing of the clock by the
  cell cycle that might have been expected to impair clock function. 

The \textit{S. elongatus} clock combines a negative transcriptional feedback oscillator (the
  transcription-translation cycle, or TTC) with a core
  phosphorylation-based post-translational oscillator (the protein
  phosphorylation cycle, or PPC).
  Remarkably, the PPC can be reconstituted \textit{in vitro} with
purified proteins \cite{Nakajima2005}, allowing detailed study of the
mechanisms behind its oscillation.  A number of studies have begun to
converge on the view that the PPC works by synchronizing the intrinsic
phosphorylation cycles of individual KaiC hexamers, primarily through
phosphorylation-dependent sequestration of KaiA by KaiC
\cite{Kageyama2006,Clodong:2007eq,VanZon2007,Rust2007,Brettschneider2010,Qin2010b,Lin:2014jf}.
Although many details of the TTC remain murkier, it seems clear that
the protein RpaA plays a central role, regulating the expression of
clock components in a manner that depends on the KaiC phosphorylation
state \cite{Takai2006,Taniguchi2010,Gutu2013,Markson2013}.  Depending
on light and nutrient levels, \textit{S. elongatus} can have doubling
times ranging from 6 to 72 h \cite{Teng2013}; the cell cycle
period is thus of the same order as the clock period of roughly 24 h,
opening the way for interactions between the two.  Indeed, the
circadian clock is known to gate mitosis, prohibiting cell division
during certain clock phases \cite{Mori1996,Dong2010,Yang:2010ci},
although in constant light this gating leaves both DNA replication and
cell growth essentially unchanged \cite{Mori1996}.  
 Conversely, Mori and Johnson argued that cell growth and division
don't affect the \textit{S. elongatus} circadian clock \cite{Mori:2001fn}. 
  We use
  mathematical modeling to study the unidirectional forcing of the
  clock by the cell cycle. We identify specific features of the \textit{S. elongatus} clock that tend to insulate it from entrainment by regular gene replication events. Nonetheless, we argue  that, under certain  conditions,
it  should be  possible to observe signatures of periodic
forcing of the clock by the cell cycle.  We further suggest how some of the
clock's protective mechanisms might be weakened experimentally,
leading to much stronger signatures of its coupling to the cell cycle.

Below, we first model the effects of cell growth and
division on a constitutively expressed protein. We show that gene
replication, not cell division, is the essential cell-cycle event that
influences protein concentrations and that, as long as the
constitutively expressed protein is not subject to rapid, active
degradation, its concentration varies little over the cell cycle.  In
contrast, gene replication
can dramatically affect
the behavior of a negative transcriptional feedback oscillator
(NTFO): the NTFO locks to the cell cycle over a range of cell-division
  times of many hours and shows erratic behavior outside this
  regime\cite{ThesisBosman}.
We next ask how the real cyanobacterial clock can be so apparently undisturbed by the cell cycle.
 We find that incorporating both a PPC and a TTC into the clock significantly weakens coupling to the cell cycle, 
 especially when the clock is read out by a push-pull network that is more sensitive to ratio of concentrations of 
 different phosphorylation states than to their absolute values.  
 The presence of multiple
chromosome copies has a still more striking effect: If the cell has 4
copies after division (rather than only 1),
as can often be the case in \textit{S. elongatus}, and if these are replicated one after the other~\cite{Jain2012a}, then
the dose of the clock genes changes much more gradually, and cell cycle effects are almost completely lost.  Thus,
\textit{S. elongatus} may have evolved to carry multiple, identical
chromosome copies in part to insulate its circadian clock from its DNA
replication cycles.

\section{Models and Results}
\subsection{The cell cycle's effect on a constitutively expressed gene is weak}
Before turning to the more complex case of a circadian clock, we first investigate how the concentration
of a single, constitutively expressed protein varies over a cell
cycle.  To this end, we add regular, rhythmic DNA replication and
mitosis to a simple model of protein production and dilution.

The key quantities in our description are the number
of copies $g(t)$ of the gene of interest and the cell
volume $V(t)$.  These vary periodically in time as
sketched in \fref{Figure1}A--B, with a period given by the cell
division time $T_{\mathrm{d}}$.  We assume for now that there
is only one gene copy present immediately after cell division.
This copy is replicated at some time before the next division, at
which point $g(t)$ jumps from 1 to 2.  When the cell
divides, the chromosomes are split between the daughter cells, and
$g(t)$ returns to 1. The cell volume grows exponentially:
$V(t)=V_{\mathrm{0}}\,\mathrm{exp}(\mudil t)$, with
$\mudil=\mathrm{log}(2)/T_{\mathrm{d}}$.  When $t$ reaches \td,
division occurs, and $V(t)$ drops back from $2 V_0$ to $V_0$.

The variables $g(t)$ and $V(t)$ define the gene density $G(t) \equiv
g(t)/V(t)$.  As long as noise and spatial variations are neglected,
the behavior of a biochemical network depends only on protein
\textit{concentrations}, not separately on protein numbers and
cell volume.  As a result, the system responds to the protein
synthesis rate per unit volume, proportional to $G(t)$, but not to $g(t)$ and $V(t)$ individually (\eref{DE_constCC}, below).
\fref{Figure1}C shows that $G(t)$ has only a single
discontinuity during the cell cycle, corresponding to the doubling of
$g(t)$ when the gene is copied; at cell division, both $g(t)$ and
$V(t)$ are halved, so their ratio is unchanged.  Importantly, then,
the meanfield, deterministic dynamics of a biochemical network is sensitive to the timing of DNA replication but not of cell
division. This dynamics is likewise unaffected by any gating of cell
division by the circadian clock, provided, as is the case in
\textit{S. elongatus} \cite{Mori1996,Yang:2010ci}, that this gating
does not affect DNA replication or cell growth. Similarly, regardless
of when during the division cycle the gene is copied, the time
dependence of $G(t)$ is always the same: It doubles, decays
exponentially for a time \td, then doubles again, etc.  The exact moment of gene replication affects only the
average value of $G(t)$, which can be absorbed,
for modeling purposes, into the parameter $\beta$ (\eref{DE_constCC},
below).  For simplicity, we thus always assume that the gene is
replicated exactly at $t = T_{\mathrm{d}}/2$.

Given the gene density $G(t)$, the concentration $C(t)$ of a
constitutively expressed protein evolves as:
\begin{equation}
 \frac{dC(t)}{dt} = \beta G(t) - \mudil C(t).
 \elabel{DE_constCC}
\end{equation}
Here, proteins are expressed at a rate
$\beta$ per gene copy and diluted by cell growth at a rate
$\mudil = \log(2)/T$.
We thus assume that, as is true for many bacterial proteins, the protein is not subject to active degradation~\cite{Lengeler1999}.
\fref{Figure1}D shows how $C(t)$ varies over
  the cell cycle.
Remarkably, even though the protein production rate doubles each time the gene
is replicated, the protein concentration varies by no more than a few percent:  The discrete jumps
in protein production are smoothed out by the slow protein
dilution. Thus, a protein that
is constitutively expressed and not actively degraded is little
affected by the cell cycle.

\subsection{The cell cycle strongly perturbs both the period and the amplitude of a negative transcriptional feedback oscillator}

Although the concentration of a protein that is constitutively expressed
 does not vary much over the cell cycle, oscillators are known to be
 far more sensitive to periodic driving than non-oscillatory systems \cite{Pikovsky2003}.
We thus next consider a simple model for a clock built on delayed, negative transcriptional feedback (\fref{Figure1}E).
The model consists of a single variable, $C(t)$, describing the
concentration of proteins that inhibit their own production:
\begin{equation}
 \frac{dC(t)}{dt} = \beta\,\tilde{G}(t)\,\frac{K_{\mathrm{c}}^n}{K_{\mathrm{c}}^n+C(t-\Delta)^n} 
  - \mu_{\mathrm{tot}} C(t).
  \elabel{DE_NFOCC}
\end{equation}
We impose a fixed delay $\Delta$
between the initiation of transcription and the appearance of
functional proteins.  
Therefore, protein production at time $t$ is proportional to the gene
copy number $g(t-\Delta)$ at time $t-\Delta$. These proteins `arrive' in the cell
volume $V (t)$ at time $t$.The
protein synthesis rate per unit volume at time $t$ is thus proportional to the \textit{protein production density}
$\tilde{G}(t) \equiv g(t-\Delta)/V(t)$.  $\tilde{G}(t)$ is a generalization of the gene density $G(t)$ of the preceding section to
the case with a delay $\Delta$ and parametrizes the periodic forcing of the NTFO by gene replication. Proteins disappear with a total rate
$\mu_{\mathrm{tot}}=\mudil+\mu_{\mathrm{act}}$, where
as before $\mudil$ describes dilution due to cell growth, and
$\mu_{\mathrm{act}}$ describes possible active
degradation.  Including both terms allows us to vary the
doubling time \td\ while holding $\mu_{\rm tot}$ constant and hence, in our
simulations, to distinguish the trivial influence of the cell cycle on
the clock through the dilution rate $\mudil$ from other
effects.

We next define the peak-to-peak time \tpeak\ as the time between successive
peaks in $C(t)$ (see \fref{Figure2} and \textit{Supporting Information} [SI]); \tpeak\ reduces to the period of the
circadian clock when oscillations are regular but remains defined when
 the cell cycle induces more erratic behavior.  In
\fref{Figure2}A we plot the average peak-to-peak time $\avg{\tpeak}$
for a range of division times \td\ at fixed $\mu_{\rm tot}$.

As expected from the general theory of driven oscillators \cite{Pikovsky2003}, the curve
shows two striking features: First, around division times which are
fractions or multiples of the clock's intrinsic period of 24 h,
the cell cycle determines the period of the clock.
Especially around $\td=24$ and 48 h, the average peak to peak time
is directly proportional to \td.  At $\td = 24$ h (1:1 locking), $\avg{\tpeak} = \td$, and the
amplitude of each clock oscillation cycle is the same
(\fref{Figure2}B).  At $\td = 48$ h (2:1 locking), however,
$\avg{\tpeak} = \td/2$, and two full clock cycles are required to make up a single division time.  Because
  these two cycles occur at different gene densities, successive peaks
  in the trace of $C(t)$ have alternately large and small
  amplitudes.

  Second, the standard deviation of \tpeak\ becomes very large
  just outside the locking regions.  \fref{Figure2}C shows that
  this variability in the phase of $C(t)$ is accompanied by
    substantial fluctuations in the amplitude for $\td=27$ h.
  Because the difference between \td\ and the intrinsic clock period
  is just too large to allow stable locking, the clock constantly
  tries to lock to the cell cycle, but slips from time to time. As a
  result, the cell cycle dramatically disrupts the clock. In the
    {\it SI} we show that both of these effects survive the
    introduction of intrinsic noise in chemical reactions and of
    stochasticity in the timing of DNA replication (Figs. S1--2;
    see also Fig. S6).  \fref{Figure3} qualitatively explains how locking
    arises in the NTFO.

\subsection{A phosphorylation cycle makes the clock more robust
  against a time-varying gene density}
\label{sec:PPC}
To study how a more realistic clock can become resilient to
variability in the gene density, we turn to the \textit{S. elongatus}
circadian clock, and more specifically to the model of Zwicker
\textit{et al.}  \cite{VanZon2007,Zwicker2010} (\fref{Figure1}F). This
model provides a detailed description of the clock, including the
synchronization of the phosphorylation state of different KaiC
hexamers via KaiA sequestration and the coupling of the PPC
oscillator to the TTC via RpaA. It represents KaiC as a hexamer
but does not explicitly take into account that each KaiC monomer has
two distinct phosphorylation sites \cite{Rust2007,Nishiwaki2007}. In
the {\em SI Text} we show that a model based on that of Rust {\it et
  al.} \cite{Rust2007}, which describes KaiC at the level of monomers
with two phosphorylation sites, gives similar results. We thus
expect that still more elaborate models of the PPC, which include
hexameric KaiC with two phosphorylation sites per monomer
\cite{Lin:2014jf}, will lead to similar results.  To include gene
replication, we modify the model of \cite{Zwicker2010} so that the
delayed negative feedback on KaiC production is modulated by a
regularly oscillating protein production density $\tilde{G}(t)$ (see
{\it SI Text}). We follow both the total KaiC concentration
$C_{\mathrm{tot}}(t)$ and the KaiC phosphorylation fraction
$p(t)=\sum_{n=1}^{6} n C_n(t)/(6\,C_{\mathrm{tot}}(t))$, where $C_n$
is the concentration of $n$-fold phosphorylated KaiC hexamers.

\fref{Figure4}A shows that a model with a PPC coupled to a TTC has a smaller locking window than an NTFO and 
lacks the large deviations in \tpeak\ just outside the locking region. 
The \textit{S. elongatus} clock is hence more robust to gene replication
than one based only on negative transcriptional feedback.

\subsection{Clock readout through an RpaA-based push-pull network filters out cell-cycle-dependent variations in protein concentrations}
\label{sec:push-pull}
Although the variance of \tpeak\ outside of the locking region is relatively small for the combined TTC-PPC model, \fref{Figure4}B shows that
  $C_{\rm tot}(t)$ exhibits strong amplitude fluctuations,
mirroring those observed for the NTFO (\fref{Figure2}). The phosphorylation fraction
$p(t)$, in contrast, is far more resilient, suggesting that the clock encodes temporal information more reliably in $p(t)$ than in $C_{\mathrm{tot}}(t)$.  
Intriguingly, the RpaA-centered push-pull network that
transmits this timing signal to downstream genes \cite{Goldbeter1981,Takai2006,Taniguchi2010,Gutu2013,Markson2013} in
fact responds primarily to $p(t)$: Because the rates of RpaA
phosphorylation and dephosphorylation are controlled by different
KaiC phosphoforms, variations in $C_{\rm tot}$ at fixed $p$ change
both rates together, leaving the fraction of phosphorylated
RpaA largely unaffected. In contrast, changes in $p$ shift the
balance between the two opposing reactions and so modify the
RpaA phosphorylation fraction (Fig. S4 and {\it SI text}). Thus,
not only is the basic PPC-based timekeeping mechanism insulated
from variations in protein synthesis, but the readout
mechanism selectively follows this more robust signal.

\subsection{Multiple chromosome copies weaken the cell cycle's influence on the clock}
\label{sec:Multiple_Copies}
While the PPC reduces gene replication's effect on the clock, it
  does not eliminate it entirely.  What other mechanisms might explain
the observed resistance of the {\it S. elongatus} clock to cell-cycle locking?  It is known that {\it S. elongatus} has
multiple, identical copies of its chromosome
\cite{Binder1990,Griese2011,Jain2012a,Chen2012}. These
are not duplicated simultaneously, but rather one at a time, so that
DNA replication occurs at a roughly constant rate throughout the cell
cycle; furthermore, the timing of chromosome duplication appears to be
independent of the phase of the clock
\cite{Binder1990,Mori1996,Watanabe2012,Chen2012,Jain2012a}. Motivated
by this observation, we consider a situation in which a cell starts
with $N$ chromosomes after division and let $g(t)$ rise to $2 N$ in
$N$ evenly spaced steps (\fref{Figure5}A). \fref{Figure5}B shows the corresponding gene
density $G(t)$. Clearly, for higher $N$, the
gene-copy number $g(t)$ increases more gradually, and hence the discrete jumps in $G(t)$ are
considerably smaller.  The effect on the clock is dramatic: The
locking regions almost disappear and the standard deviation in \tpeak\
becomes negligible (\fref{Figure5}C).  Multiple chromosomes
similarly make the NTFO much less susceptible to gene replication, but in
the absence of the PPC cell-cycle effects are not blocked
so completely (Fig.~S5). Fig. S6 summarizes the
  combined effects of chromosome number and variability in gene
  replication time on our clock models.  Importantly, at small $N$
gene replication always significantly affects the clock, through
  either phase locking or high variability in \tpeak.

\section{Discussion}
Given the pleiotropic roles of both the cell cycle and the
 circadian clock, it is natural to
ask whether they also influence each other.  Our central observation
is that such influence need not involve specific interactions between
the core genes or proteins of the two systems
\cite{Johnson2010,Feillet2014,Bieler2014}; rather, the simple fact
that the number of cellular copies of a given gene necessarily experiences
discrete jumps during DNA replication (\fref{Figure1}) implies that
 clocks must in general feel a periodic driving from the cell
cycle \cite{ThesisBosman}. Whereas some genetic circuits can simply average over this
time-varying input, oscillators---including biological clocks---are
known to be especially sensitive to rhythmic forcing.  Indeed,
an NTFO either locks to the cell cycle or shows
erratic oscillations for a range of doubling times
\td~(\fref{Figure2}), losing its ability to function as a clock in either case.

In light of this strong and detrimental coupling between the cell
  cycle and a simple
  transcriptional clock, it is all the more
  striking that the \textit{S. elongatus} clock is so stable.
Our analysis highlights two features of the cyanobacterial clock that are
  predicted to allow the necessary decoupling from the cell cycle.
 First, a time-varying gene dosage
influences a clock with an autonomous post-translational oscillator
less than it does a purely transcriptional clock; even {\em within}
the combined TTC-PPC, the oscillations of the KaiC phosphorylation
fraction $p(t)$ are less affected by periodic gene replication than
are those of the total KaiC concentration $C_{\rm tot}(t)$
(\fref{Figure4}, S2C).  Strikingly, the RpaA-based push-pull network
that communicates the clock state to the rest of the cell
responds to $p$ while ignoring the more strongly fluctuating
$C_{\rm tot}$ (somewhat in the spirit of mechanisms that improve
the  robustness of bacterial chemotaxis to gene expression noise
\cite{Kollmann:2005fz}). This filtering function of the
push-pull architecture could help explain why the
\textit{S. elongatus} clock has a relatively complex output mechanism
requiring both CikA and SasA rather than a simpler linear
design~\cite{Shultzaberger2014}.
%More generally, these results
%underscore the importance of the PPC in providing robustness to the
%clock \cite{Johnson2008c,Zwicker2010}.

The second feature of the {\it S. elongatus} clock that we predict mitigates
perturbations from the cell cycle is the presence of multiple,
identical, asynchronously replicating chromosome copies
\cite{Binder1990,Watanabe2012,Chen2012,Jain2012a}.  This reduces the importance of each individual gene replication
event: Rather than seeing a single doubling of the number of gene
copies each cell cycle, a cell with many chromosomes instead sees a
number of smaller jumps that it can more easily ignore (\fref{Figure5}). This
adaptation may thus have evolved in part to protect the
\textit{S. elongatus} clock from cell cycle effects.

Whereas we have argued that the cell cycle generically affects any
transcriptional clock, no comparably general mechanisms exist in the
other direction.  Moreover, though in many eukaryotic systems the
clock is known to regulate key cell-cycle genes
\cite{Johnson2010,Masri2013,Sotak2014,Mori1996,Matsuo2003,Nagoshi2004,Dong2010,Yang:2010ci,Gerard2012},
no similar, specific connections have yet been characterized in
\textit{S. elongatus}.  In particular, clock-dependent cell-cycle
gating \cite{Mori1996}, because it acts on cell division but not on
growth or DNA replication, does not allow the clock to block the
discrete gene replication events that underlie the driving. Nonetheless, since the majority of
\textit{S. elongatus} genes shows some degree of clock-dependent
expression \cite{Ito2009}, it is possible that the cyanobacterium's
clock does regulate its cell cycle in some as yet undiscovered way.
Any such coupling would however have to be weak enough to be
consistent with the observation that the rhythm of DNA replication
does not depend on clock phase
\cite{Binder1990,Mori1996,Watanabe2012,Chen2012,Jain2012a}.  Because
phase locking between two oscillators has strong similarities to the
locking of a single oscillator to periodic driving
\cite{Pikovsky2003}, most of our qualitative conclusions would remain
unchanged in this case.

  To isolate the behavior of the core, autonomous circadian
  oscillator, studies in the lab are typically performed at
  constant light levels.  In keeping with this tradition, we have
  limited ourselves here to models of free-running clocks, without any diurnal environmental variation.
   In nature, however, the circadian clock is
  exposed to many additional entrainment signals, most notably the 24 h
  light-dark cycle. In fact, the environmental and
  cell cycle entrainment signals are intricately intertwined,
  because DNA replication and the synthesis of most
  proteins, including clock components, come to a standstill in the dark in a
  clock-independent fashion \cite{Tomita2005,Watanabe2012}.  
  We leave the effects of this complex interplay for future work.

 Although we have focused on interactions between the cell cycle
  and the clock in \textit{S. elongatus}, the basic
  idea that periodic gene replications must influence biological oscillators
  is more general and should apply to a wide range of
  prokaryotic and eukaryotic species.  Indeed, cell-cycle-dependent
  changes in gene copy number have clearly observable effects on gene
  expression in eukaryotic cells \cite{Zopf:2013bb}, and recent experiments in
  cultured metazoan cells strongly suggest that the cell cycle exerts
  a considerable influence on the circadian clock, generally leading
  to phase locking of the two oscillators
  \cite{Feillet2014,Bieler2014}. Other generic forms of driving from the cell cycle
  may also play a role here: for example, in contrast to prokaryotes, eukaryotes typically shut
  down transcription around mitosis,
  thereby introducing another source of periodic, cell-cycle dependent variation in
  protein synthesis \cite{Johnson2010,Feillet2014,Bieler2014}. 
  Our analysis thus highlights an important constraint on the design of circadian clocks in organisms from bacteria to humans.

  Further, there is no reason for the effects of regular,
  discrete gene replications to be limited to circadian clocks; they
  should be observable in any cellular oscillator that depends on
  transcription and has a period on the same order as that of the cell
  cycle.  Thus, our results may be relevant to phenomena like coupling
  between the cell cycle and the segmentation clock in vertebrate
  development \cite{delaune2012}.  Similarly, in the
    {\it SI} (Figs. S7--S8) we show that two
    well-known synthetic circuits \cite{Elowitz2000,Stricker2008} can also lock to the
    cell cycle, and that the strength of locking depends sensitively
    on the oscillator architecture.

  Since we have argued that \textit{S. elongatus} possesses particular
  adaptations that decouple its circadian clock from the cell cycle,
  the most obvious experimental test of our ideas would be to observe
  the consequences of blocking or removing these features. Several strains already exist that might allow just such experiments. Mutants of \textit{S. elongatus} are known with
  significantly fewer chromosomes per cell than the wildtype \cite{bird1998}; moreover, in some other \textit{Synechococcus} strains, cells are always monoploid~\cite{Griese2011}.  We find that in
  cells where the number of chromosomes goes from 1 to 2 over the
  course of a single division cycle, it should be possible to observe clear
  signatures of driving by the cell cycle in plots of KaiC's
  abundance---but not its phosphorylation level---as a function of
  time (\fref{Figure4}).
  We predict that this effect will be further strengthened if the PPC
  is removed entirely.  It is well-established that this can be
  accomplished by hyper-phosphorylating KaiC 
  \cite{Kitayama2008,Qin2010}. In all cases, one could study forcing by
  the cell cycle at a variety of different doubling times. We suggest, however, that a doubling time near
  48 hours offers a particularly unambiguous signature of the cell
  cycle's influence: The KaiC abundance as a function of time should
  then rise and fall every 24 hours, with successive peaks strictly
  alternating between higher and lower levels (\fref{Figure4}C).

\subsection*{Acknowledgments}
  We thank Jeroen van Zon for a critical reading of the manuscript. This work was supported in part by FOM, 
  which is financially supported by the Nederlandse Organisatie voor Wetenschappelijk Onderzoek
  (JP, MB and PRtW), and by NSF Grant DMR-1056456 (DKL).

\bibliographystyle{unsrt}
%\bibliography{library}

\nocite{Bogan2001}

\begin{figure}[bth!]
\begin{center}
\vspace{-30pt}
\includegraphics[]{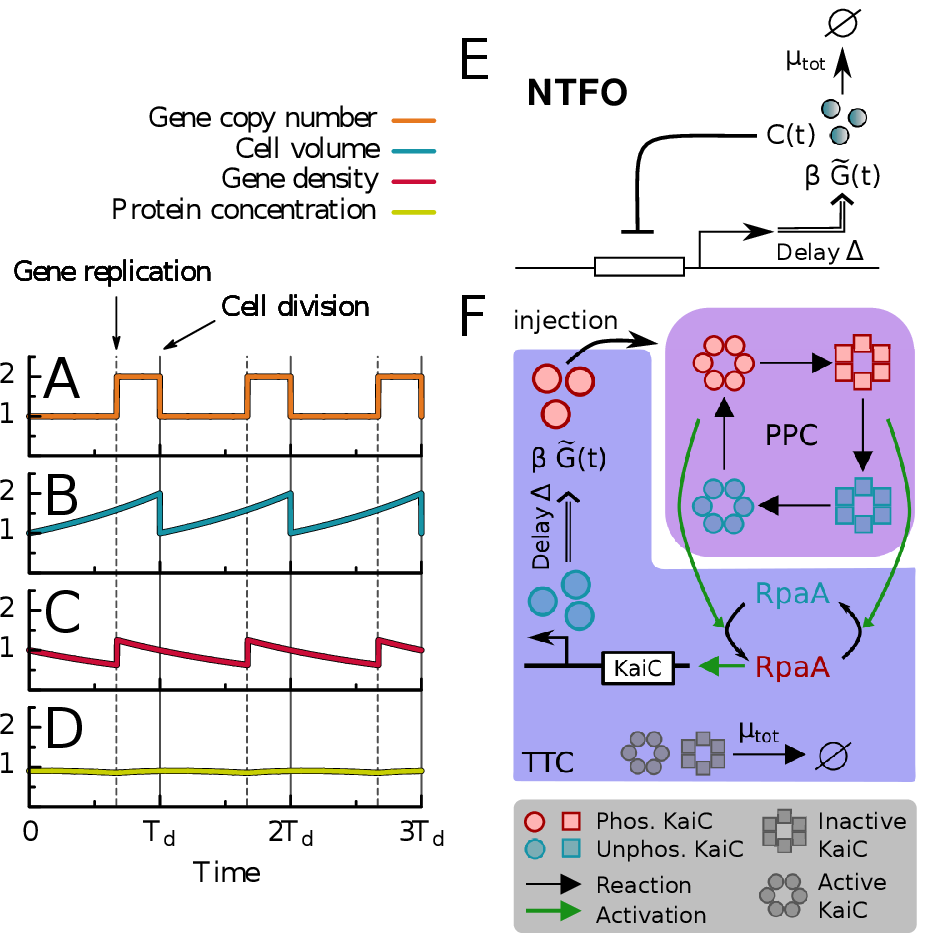} 
\end{center}
\caption{\flabel{Figure1} DNA replication but not cell division
    affects average expression levels; for a protein that is
    constitutively expressed and decays by dilution only, the effect is
    small.
  Schematic time courses
  of the gene copy number $g(t)$ (A), the cell volume $V(t)$ (B), the
  gene density, $G(t)=g(t)/V(t)$ (C), and the concentration
  $C(t)$ of a constitutively expressed protein that decays only
  by dilution (D).  Time in units of the cell division
  time $T_{\mathrm{d}}$; vertical axes, arbitrary units.
  The gene density (C) has a discontinuity when the gene is
  replicated (vertical dotted lines) but not at cell
  division (vertical solid lines), when both $g(t)$ and $V(t)$ are
  halved.  Even though the protein synthesis rate doubles when the
  gene is replicated, the maximum deviation of $C(t)$ from its time average is less than 4\% (D).  
  (E) The NTFO model: A protein with concentration $C(t)$ represses its own transcription with a delay $\Delta$.  
  (F)  Zwicker \protect\cite{Zwicker2010} model for coupled phosphorylation (PPC, purple background) and transcription-translation (TTC, blue background) cycles. 
KaiC hexamers switch
between an active conformational state (circles) in which their phosphorylation
level tends to rise and an inactive state (squares) in which it tends to fall.
Active KaiC activates RpaA and inactive KaiC inactivates RpaA; active
RpaA (red) activates \textit{kaiBC} expression, leading (after a delay) to the injection
of fully phosphorylated KaiC (pink) into the PPC.}
\end{figure}

\begin{figure}[!bth]
\begin{center}
\vspace{-15pt}
\includegraphics[]{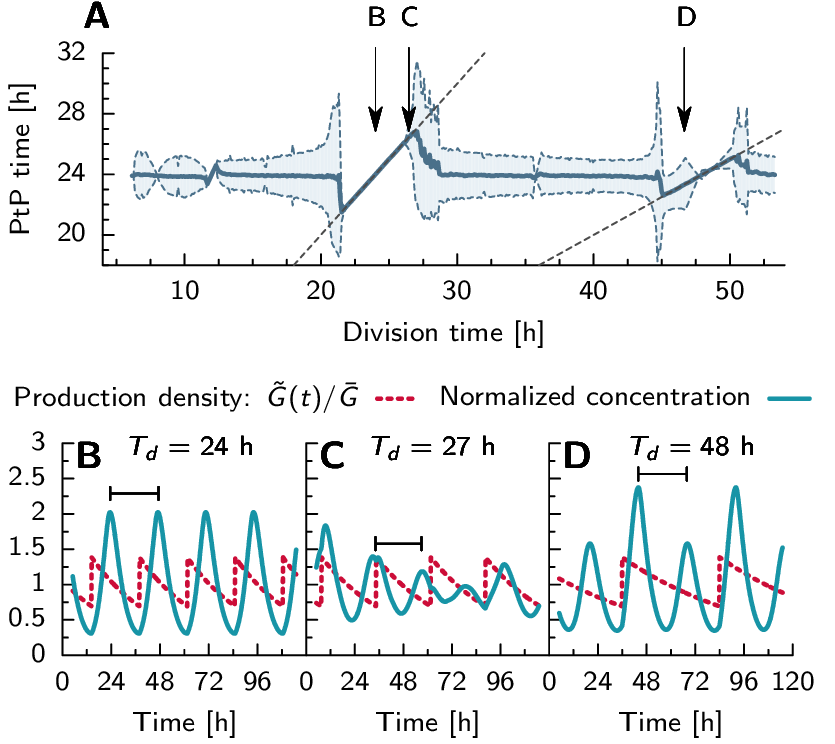}
\end{center}
\caption{\flabel{Figure2} Periodic gene replication dramatically affects a negative
    transcriptional feedback oscillator (NTFO).  (A) The average
  peak-to-peak time $\avg{\tpeak}$ (solid curve) versus the
  cell division time \td\ at fixed $\mu_\text{tot}$ and $\beta$.  The shaded
  region shows the standard deviation of the peak-to-peak times (see
  \textit{SI text}). Dashed lines indicate regions where the clock
  locks to the cell cycle with periods in a 1:1 (left) or 2:1 (right)
  ratio.  (Smaller locking regions around $\td = 6, 12, \text{and}
  \, 36$ h are not marked.)
(B--D)
    Protein concentration $C(t)$ (blue solid line) and
    the protein production density $\tilde{G}(t) = g(t-\Delta)/V(t)$
    (red dashed line) for the values of $T_\text{d}$ indicated by the arrows in (A); horizontal brackets in (B--D) illustrate the definition of the peak-to-peak time \tpeak.  
  At $\td = 24$ h (B), the clock locks firmly to the cell cycle.  For
  $\td=27$ h (C), the cell-cycle period
 is just too large for locking; as a result, the cell cycle dramatically disrupts the clock, leading to a
  large standard deviation of \tpeak\ (see panel A).  At $\td=48$ h (D), two oscillation cycles of the
  NTFO fit exactly in one division time. The larger
  amplitude oscillation cycle corresponds to cell cycle phases where $\tilde{G}(t)$ is
  higher and the smaller amplitude to phases where
  $\tilde{G}(t)$ is lower. Similar results are obtained upon
    varying \td\ at constant $\mu_{\rm act}$ (Fig. S3).}
\end{figure}
    
\begin{figure}[b]
\begin{center}
\vspace{-15pt}
 \includegraphics[]{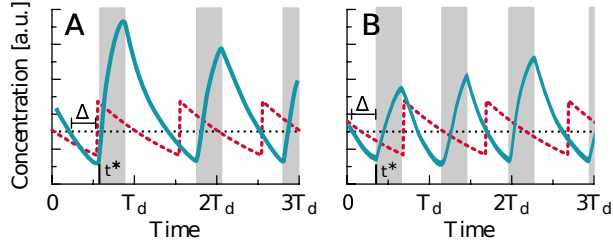} 
\end{center}
\caption{\flabel{Figure3} Locking mechanism for the NTFO.  Shown are time courses of the production density $\tilde{G}(t) =
  g(t-\Delta) / V(t)$ (dashed red lines) and the protein
  concentration $C(t)$ (solid blue lines). For clarity, we consider the
    limit $n\to\infty$, in which the Hill function describing
    autoregulation (\eref{DE_NFOCC}) reduces to a step function with repression threshold $K_{\mathrm{c}}$, denoted by the
    dotted horizontal line. Shaded regions indicate times when
    $C(t)$ is rising. The panels correspond to two different
  initial phase differences between the NTFO and the cell cycle. In
  each case, when $C(t)$ drops below $K_{\mathrm{c}}$ at time
  $t^*-\Delta$, protein production starts, but because of the delay
  $\Delta$, new molecules are injected into the system
  only at time $t^*$.  (A) The gene has replicated just before
  $t^*-\Delta$, and $\tilde{G}(t^*)$ is hence large, yielding a
    large amplitude for the next
  NTFO cycle.  Because the rate of protein
  decay is independent of $\tilde{G}(t)$, the period of the NTFO cycle is correspondingly long. The subsequent NTFO
  cycle thus begins at smaller $\tilde{G}(t^*)$, causing it to have a smaller amplitude and a shorter period. (B) The gene has not yet replicated at time
  $t^*-\Delta$, and $\tilde{G}(t^*)$ is therefore
  low; consequently, the amplitude and period of the next NTFO cycle are small. The beginning of the subsequent cycle is then shifted
  towards higher $\tilde{G}(t^*)$, increasing its period. In both cases, the result is
  that, after a few cell cycles, the period of the NTFO
  oscillation approaches that of the cell cycle, yielding
    stable 1:1 locking where the two oscillators have a well-defined phase
    relation. The largest amplitude and thus longest possible clock period arise when the protein synthesis phase (grey bar) coincides with the maximal $\tilde{G}(t^*)$; if \td\ increases beyond this maximal period, locking cannot occur.  An analogous loss of locking occurs if \td\ decreases below the minimal possible clock period. In either case, the clock shows erratic behavior until \td\ approaches values where 1:2 or 2:1 locking is possible. }
\end{figure}

\begin{figure}[!hbt]
\begin{center}
\vspace{-15pt}
  \includegraphics[]{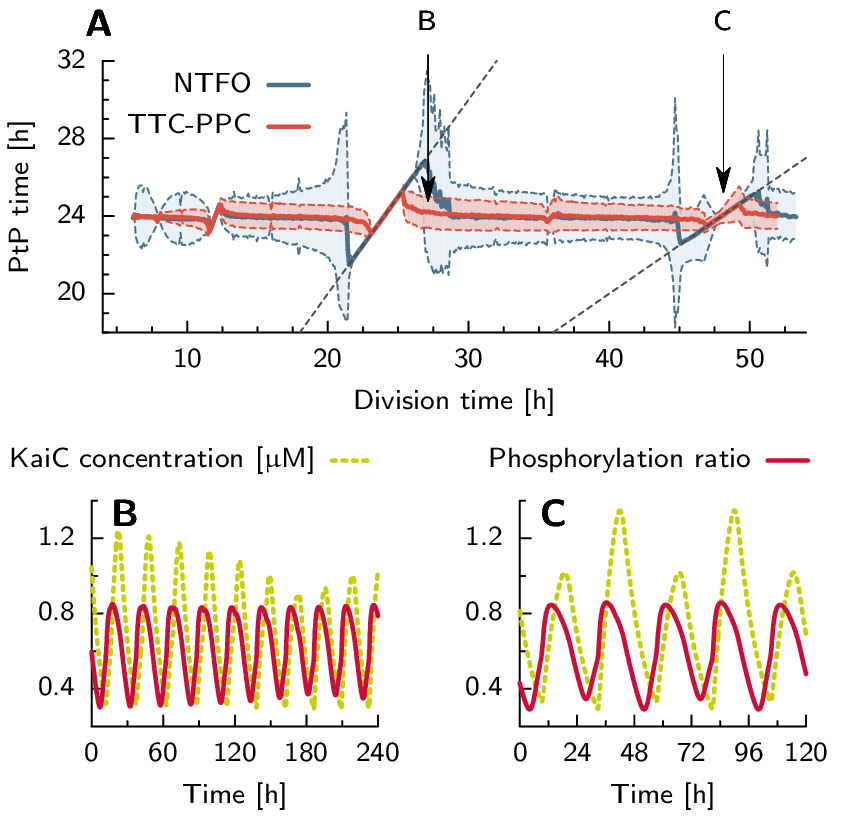} 
\end{center}
\caption{\flabel{Figure4} A clock with interlocked phosphorylation and transcriptional cycles is more robust against
    perturbations from periodic gene replication.  (A) The average
  peak-to-peak times $\langle \tpeak\ \rangle$ of the phosphorylation level $p(t)$ of the
    coupled PPC-TTC model of the Kai system \cite{Zwicker2010} (red
  solid curve) and of $C(t)$ of the NTFO (solid blue curve,
  same as \fref{Figure2}A), as a function of the cell division time
  \td.  The shaded regions show the standard deviation of \tpeak.  Both the widths of the locking regions and the
  standard deviations of the peak-to-peak time outside the
  locking regions are smaller for $p(t)$ of the Kai system than
    for $C(t)$ of the NTFO.  Arrows indicate division times for which
    we show time traces in (B,C).  (B) The total KaiC
  concentration $C_{\rm tot}(t)$ (dashed line) and $p(t)$ (solid line)
  at  $\td=26$ h.  Though the amplitude of $C_{\rm tot}(t)$ is
    strongly affected by gene replication, the amplitude of $p(t)$ is
    nearly constant.  (C) Plots of $p(t)$ and $C_{\rm tot}(t)$ at $\td=48$ h, where the amplitude of
  $C_{\mathrm{tot}}(t)$ alternates between a low and a high value depending on
  the gene copy number in the cell. In contrast, $p(t)$ is
    almost unaffected by gene replication.  }
\end{figure}

\begin{figure}[b]
\begin{center}
\vspace{-15pt}
  \includegraphics[]{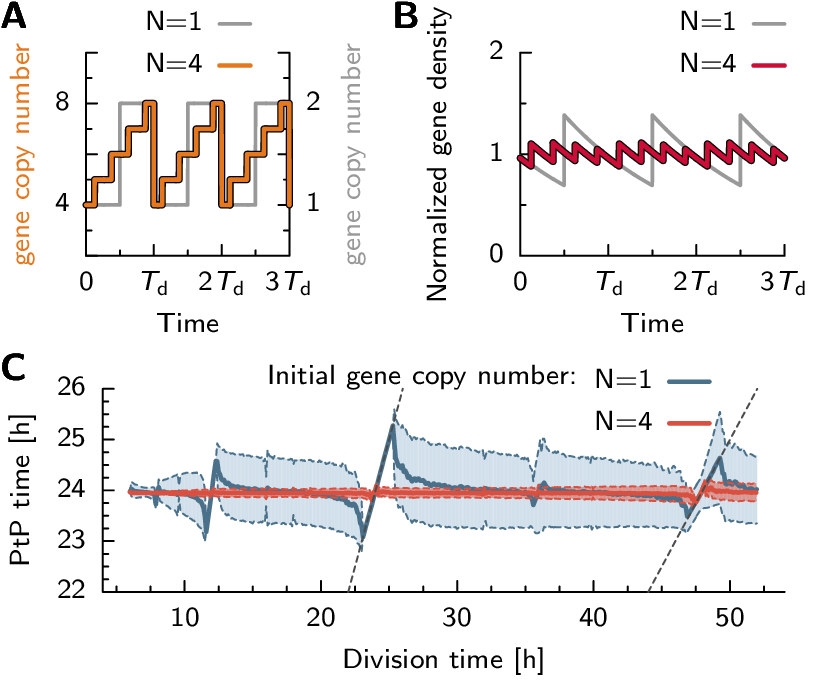} 
\end{center}
\caption{\flabel{Figure5} A higher average gene copy number strongly
  reduces the effect of the cell cycle on the circadian clock.  (A)
  Time course of the gene copy number $g(t)$ for initial gene copy
  numbers $N=4$ (thick
  curve, left axis) and $N=1$ (thin curve, right axis); time in units of cell cycle time \td.  The increase in $g(t)$ is
  more gradual for $N = 4$ than for $N = 1$.  (B) The gene
  density $G(t) = g(t) / V(t)$, normalized to its time average, for $N=4$ (thick curve) and $N=1$ (thin
  curve). At a higher gene copy number, the deviations from the
  average gene density become smaller.  (C) The average peak-to-peak
  time $\langle \tpeak \rangle$ of the phosphorylation fraction $p(t)$ of the PPC-TTC model of
  the Kai system \cite{Zwicker2010}, for initial gene copy numbers
  $N=1$ (solid blue curve, same as \fref{Figure4}A) and $N=4$ (solid
  red curve) versus cell division time \td.  (Note the
  $y$-axis range is smaller than in ~\fref{Figure2}A and
  \fref{Figure4}A.)  For the higher gene copy number, the locking
  regions have almost disappeared and the standard deviation in the
  peak-to-peak times is very small. For time traces, see Fig. S5.}
\end{figure}

\end{document}